# Multi-Institutional Audit of FLASH and Conventional Dosimetry with a 3D-Printed Anatomically Realistic Mouse Phantom
## Running Title: FLASH Dosimetric Audit with a 3D Printed Mouse Phantom

M Ramish Ashraf*[1], Stavros Melemenidis*[1,] Kevin Liu[3], Veljko Grilj[4], Jeannette Jansen[5], Brett Velasquez[3], Luke Connell[3], Joseph B Schulz[1], Claude Bailat[4], Aaron Libed[1], Rakesh Manjappa[1,] Suparna Dutt[1], Luis Soto[1], Brianna Lau[1], Aaron Garza[1], William Larsen[1], Lawrie Skinner[1], Amy S Yu[1], Murat Surucu[1], Edward E Graves[1,6], Peter G Maxim [2], Stephen F. Kry[3], Marie-Catherine Vozenin[#5], Emil Schüler[#3], Billy W Loo Jr[#1,6]

[1] Department of Radiation Oncology, Stanford University School of Medicine, Stanford, CA, 94305

[2] Department of Radiation Oncology, University of California, Irvine, CA, 92697

[3] Department of Radiation Physics, Division of Radiation Oncology, The University of Texas MD Anderson Cancer Center, Houston TX, 77030

[4] Institute of Radiation Physics, Lausanne University Hospital and University of Lausanne, Switzerland

[5] CHUV- Radiation-Oncology Laboratory

[6]Stanford Cancer Institute, Stanford University School of Medicine, Stanford CA 94305

*Co-first authors

[#] Co-senior/co-corresponding authors: bwloo@stanford.edu, eschueler@mdanderson.org, marie-catherine.vozenin@chuv.ch

Acknowledgements: We thank Christine F. Wogan, MS, ELS, of MD Anderson's Division of Radiation Oncology, for editorial contributions to this article. This study was supported by funding from NIH grants P01CA244091 (EEG, PGM, BWL, MCV), R01CA26667 (ES, PGM, EEG, BWL), AAPM-ASTRO Trainee Seed Grant (MRA), and by UTHealth Innovation for Cancer Prevention Research Training Program Pre-Doctoral Fellowship (Cancer Prevention and Research Institute of Texas grant #RP210042 (KL)). We also gratefully acknowledge philanthropic donors to the Department of Radiation Oncology at Stanford University School of Medicine. The content is solely the responsibility of the authors and does not necessarily represent the official views of the National Institutes of Health nor of the Cancer Prevention and Research Institute of Texas.

Disclosures: BWL has received research support from Varian Medical Systems, is a co-founder and board member of TibaRay, is a consultant on a clinical trial steering committee for Beigene and has received lecture honoraria from Mevion.

**Background:** Means for dosimetric credentialing of institutions for FLASH preclinical research analogous to the credentialing done for cooperative group clinical trials of radiation therapy are currently lacking.

**Purpose**: We conducted a multi-institutional audit of dosimetric variability between FLASH and conventional dose rate (CONV) electron irradiations by using an anatomically realistic 3D-printed mouse phantom.

**Methods:** A CT scan of a live mouse was used to create a 3D model of bony anatomy, lungs, and soft tissue. A dual-nozzle 3D printer was used to print the mouse phantom using acrylonitrile butadiene styrene (~1.02 g/cm$^3$) and polylactic acid (~1.24 g/cm$^3$) simultaneously to simulate soft tissue and bone densities, respectively. The lungs were printed separately using lightweight polylactic acid (~0.64 g/cm$^3$). Hounsfield units (HU) and densities were compared with the reference CT scan of the live mouse. Print-to-print reproducibility of the phantom was assessed by CT imaging and electron beam irradiation. Three institutions were each provided a phantom, and each institution performed two replicates of irradiations at selected mouse anatomic regions, with five film irradiations per anatomic region, per replicate. The average dose difference between FLASH and CONV dose distributions and deviation from the prescribed dose were measured with radiochromic film.

**Results:** Relative to the reference CT scan of the live mouse, CT scans of the phantom demonstrated mass density differences of 0.1 g/cm$^3$ for bone, 0.12 g/cm$^3$ for lung, and 0.03 g/cm$^3$ for soft tissue regions. Between phantoms, the difference in HU for soft tissue and bone was <10 HU from print to print. The lung showed the greatest variation (54 HU), but this had minimal influence on the dose distribution, with <0.5% dose differences between printed phantoms. For all irradiations (across institutions and replicates), the mean difference between FLASH and CONV doses was 2.8%. The mean difference from the prescribed dose for all irradiations was 3.1% for CONV and 4.6% for FLASH. The mean difference between FLASH and CONV from the first replicate to the second decreased from 4.3% to 1.2%, and the mean difference from the prescribed dose decreased from 3.6% to 2.5% for CONV and 6.4% to 2.7% for FLASH.

**Conclusions:** This audit using a 3D printed mouse phantom revealed that FLASH and CONV electron irradiation can be delivered consistently between participating institutions. The framework presented is promising for credentialing of multi-institutional studies of FLASH preclinical research to maximize the reproducibility of biological findings.

## 1. Introduction

Ultra-high dose rates (>40 Gy/s) have been shown to spare normal tissue relative to conventional dose rate (CONV) radiotherapy without compromising antitumor efficacy[1–8]. The biological mechanisms underlying this phenomenon, termed the "FLASH effect," remain unclear. Although further preclinical studies are needed to establish the optimal beam parameters, a vast body of evidence has already been published with different radiation modalities and animal models[4,9]. Different outcomes have been reported with regard to the FLASH effect across different institutions, even though similar experimental setups were used[10,11]. However, recently Zayas et al[12] showed that if certain beam parameters are matched, the FLASH effect can be reproducible across institutions. With funding of up to US $15M from the US National Institutes of Health invested in FLASH-related research[13] and human trials already underway[14–16], it is essential that dosimetric audit processes be implemented to reduce technical variability from study to study.

Preclinical dosimetry in general has been problematic, which has led to reproducibility issues in the past[17,18]. Freedman et al reported that about US $28B/year is spent on preclinical studies that are not reproducible[19]. A recent (2021) review article on dosimetry for radiobiological studies by DeWerd et al[20] showed that the dose variation against the reference dose typically ranged from 12% to 42%. Daeger et al[21] conducted a comprehensive literature review of preclinical dosimetry and found that most publications on preclinical research do not report important information such as dose specification, beam energy, and irradiation setup. Preclinical dosimetry is challenging primarily because of the use of small field sizes and non-standard machines for which reference conditions compared with standardized protocols are lacking. Dosimetry under FLASH conditions is even more complex and error-prone; the issue of non-standard conditions is exacerbated as electron FLASH beamlines are often obtained by custom configuration of existing clinical linear accelerators. This is typically achieved by disabling feedback mechanisms and requires in-house–built solutions for beam control[22–25]. Lack of treatment planning software for FLASH is another major contributor to the complexity of delivering dose under FLASH conditions. Also, at FLASH dose rates most radiation detectors that are commonplace in the medical physics community are no longer appropriate owing to saturation and dose-rate dependence[26]. Moreover, FLASH and CONV dose deliveries can be made with different beam energies or mismatched

beam geometries with different source-to-surface distance (SSD)[8]. This further complicates the ability to deliver the same spatial dose distribution under both conditions.

Jorge *et al*[27] recently conducted a dosimetric audit for FLASH using a homogenous acrylic cuboid phantom with three different dosimeters: (1) thermoluminescent dosimeters, (2) alanine pellets, and (3) radiochromic film. Two institutions were evaluated with different beam parameters, and the results for the three dosimeters were generally within 3% of the expected target dose in the middle of the phantom. However, that study was performed with a homogenous phantom; because treatment planning software and Monte Carlo models are not widely available for FLASH, areas of heterogenous densities are where the greatest discrepancies can arise between delivered and prescribed doses. Moreover, most institutions calibrate dose by using a film at the entrance surface of the subject and scaling the dose measured by the entrance film to the dose at target depth by using the percentage depth dose curve acquired with homogenous solid water phantoms. This approach works for homogenous areas but risks introducing unacceptable errors, particularly for small fields and in anatomy with substantial heterogeneities such as lung.

The goals of this study were to construct and validate an anatomically realistic 3D printed mouse phantom for electron beam irradiation and perform a multi-institutional audit for FLASH using this phantom and radiochromic films. Although 3D printed phantoms have been used for preclinical dosimetric audits and imaging studies[28–33], including mail-order peer review [34], a systematic audit evaluating the consistency between and accuracy of FLASH and CONV irradiation platforms has never been done. The phantom, films, and the associated protocols were sent to three institutions (Stanford University, MD Anderson Cancer Center, and Centre Hospitalier Universitaire Vaudois [CHUV]). The institutions were instructed to deliver a target dose to the middle of the phantom by using both CONV and FLASH dose rates. Two replicates were performed by each institution. This approach could serve as a framework for dosimetric credentialing of institutions for FLASH preclinical research analogous to that done for cooperative group clinical trials of radiation therapy, which is currently lacking.

## 2. Methods and Materials

### 2.1 Film Dosimetry

In this study, we used EBT-3 Gafchromic film (Ashland, Wilmington, DE), which has previously demonstrated dose-rate independence at FLASH dose rates[35]. To ensure precise positioning within the phantom, a laser cutter (Beambox®) was used to cut the film into desired shapes. Electron collimators were designed to irradiate three fields: longer and shorter rectangular fields over the body of the mouse phantom, and one for the head region. (From here on, body irradiation to the longer field will be referred to as 'Body Long' and irradiation with the shorter field as 'Body Short'.) Films were calibrated by using a clinical 12-MeV beam to deliver known dose levels on a clinical linear accelerator to establish the relationship of the optical density (OD) with dose. The beam output was verified by using a TG-51 setup within a farmer chamber calibrated by an accredited dosimetry calibration laboratory. Films were scanned one by one in RGB mode, 48 bits, 72 dpi, without color correction (EPSON Scanner 10000XL)) and placed at the center of the scanner reproducibly by using a registration template. Only the red channel was used for dosimetry. Notably, even though triple channel dosimetry[36] can give superior results with reduced uncertainty, most artifacts and dose-independent perturbations that the triple-channel uniformity method avoids are already minimized in the current approach. For example, a template is used to ensure repeatable positioning of each piece of film on the scanner's central axis. Because the pieces of film are small and scanned near the central axes of the scanner, potential sources of error from off-axis scanning and film curling are minimized. Further, each laser-cut film has a unique piece of code engraved on it by the laser cutter and is scanned before irradiation. This serves as a unique background image for each piece of film. Net OD was calculated as:

$$\text{netOD (x,y)} = \log_{10}\left(\frac{PV\ (x,y)_{\text{bkg}}}{PV\ (x,y)_{\text{exp}}}\right)$$

where (x, y) are International Electrotechnical Commission x- and y-coordinates and indicate that pixel-wise subtraction is performed, and PV is the pixel value. The subscript *bkg* refers to the background film scanned before irradiation, and *exp* refers to the experimental film. The two images (*bkg* and *exp*) were rigidly registered by using holes in the film as landmarks. The subtraction process accounts for any inherent inhomogeneity in the active layer or local changes in background value for each piece of film. A 5th order polynomial was used for calibration. The calibration curve was validated by delivering known doses under calibration conditions and comparing them against measured dose. The delivered doses were scaled to reflect any changes in beam output as verified by an ionization chamber. A major source of uncertainty in film-based dosimetry is due to changes in netOD after exposure[37,38]. This is especially critical when film is to be used in a multi-institutional dosimetric audit because the time to return the films may vary. Ideally the experimental films should be scanned at a post-exposure time that matches as closely as possible the post-exposure time used for the calibration curve. For this purpose, calibration curves were obtained at multiple time points after exposure (24, 36, 48, 136, and 336 h), providing various calibration curves for post-exposure read-out. The calibration curve that most closely matched the post-exposure delay between irradiation and read-out of the experimental films was used[38]. Additional films sent to all institutions were to remain unirradiated, to be used as 'shipment background control' films to account for any overall changes to OD during shipments, which can occur because of variations in temperature, humidity or exposure to incidental radiation during the shipping process[39].

## 2.2 3D Printed Mouse Phantom

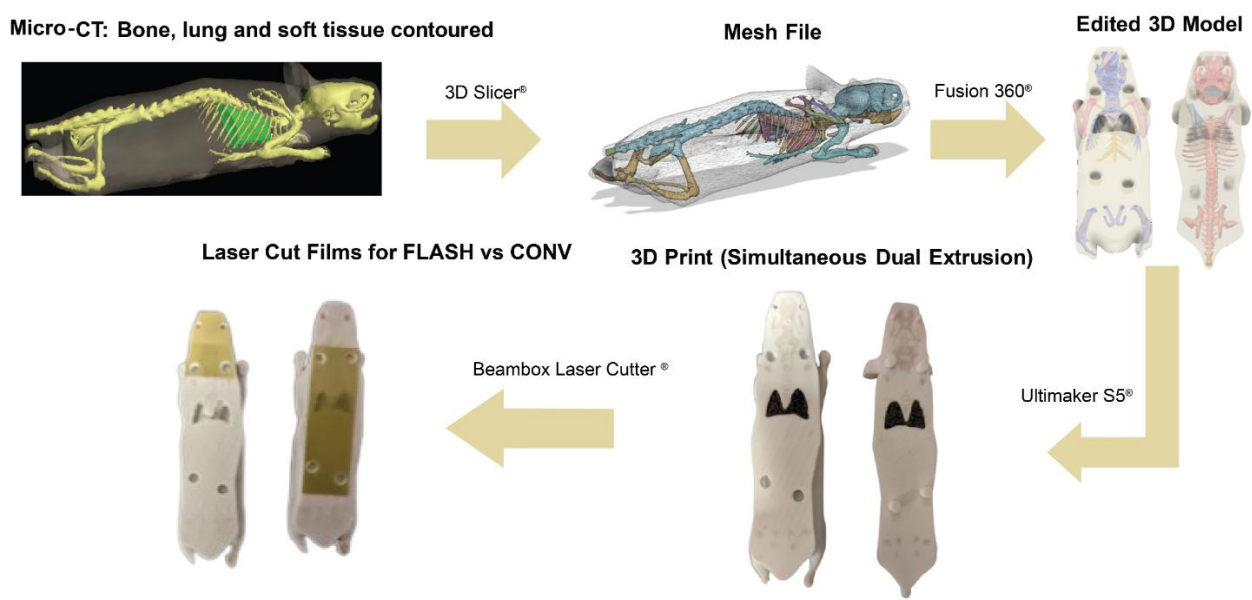

**FIGURE 1**. *Outline of the general workflow to develop the 3D printed mouse phantom. Bone, lungs, and soft tissue were contoured by using a 3D micro-CT dataset. The contours were converted to mesh files, imported into CAD software, and edited by splitting the phantom in half and including registration pegs for radiochromic film positioning. Printing parameters were optimized, and material densities were verified by weighing cubes of known volume. Mesh files were printed by using dual simultaneous extrusion with polylactic acid (PLA) and acrylonitrile butadiene styrene (ABS) printing materials to mimic bone and soft tissue, respectively. Lightweight PLA was used to mimic lungs. Radiochromic films were laser-cut to cover different anatomic sites and ensure accurate positioning within the phantom.*

The workflow for developing the 3D-printed mouse phantom is illustrated in Figure 1. All animal procedures were approved by the Stanford University Institutional Animal Care and Use Committee (IACUC), and the work in this paper followed these approved procedures throughout the study. A micro-CT scan of a mouse was obtained (Precision X-Ray Inc, Madison, CT, USA; kVp=80, 1 mA, pixel size=0.4 mm$^2$) and was exported to an Eclipse treatment planning system (Varian, Palo Alto, CA, USA), where bone, lungs and soft tissue were contoured. The contours were then converted to mesh files by using an open-source platform (3D Slicer). The mesh files were imported into Fusion 360 (Autodesk, San Rafael, CA, USA) and converted to a 3D editable object. The 3D model was modified to allow the inclusion of Gafchromic film to enable 2D dosimetry for different anatomic sites. The registration pegs for films were designed for accurate positioning in a unique orientation, leaving no ambiguity in the direction of the film during irradiation. Finally, the edited model was processed with Ultimaker Cura v.4.3.1 and printed with the Ultimaker S5 (Ultimaker, Utrect, Netherlands), a fused deposition modeling–based 3D printer. The 3D model consists of three separate sub models corresponding to the three contoured anatomic tissue types. Bone and soft tissue sub models were merged and printed together. Because both sub models originated from the same micro-CT scan, merging them in the software maintains their relative positions.  The Ultimaker S5 has two nozzles that are used together to print soft tissue and bony anatomy in a single print. Lung equivalent material was printed separately and added manually to the printed phantom.

Acrylonitrile butadiene styrene (ABS) (MatterHackers, Lake Forest, CA, USA) was used to mimic soft tissue, and polylactic acid (PLA) (MatterHackers, Lake Forest, CA, USA) was used to mimic bone density. ABS has an inherent density of ~1-1.05 g/cm$^3$, which makes it an ideal candidate for reproducing radiologic properties of soft tissue, whereas the inherent density of PLA (~1.24 g/cm$^3$) makes it suitable for printing more dense anatomic features[40]. Although materials that are more dense than PLA are available, such as nylon, vinyl and combinations of PLA and chalk/stone[41], for this study we found PLA to be a reasonable choice because only electron beams were used and because PLA is easy to print and widely available. PLA would not be an appropriate general surrogate for bone, especially for kV photon irradiation, because of its low Z. These two materials were also selected because their HU value has been shown to be more stable over time than other 3D printed materials (i.e., <8 HU over 6 weeks)[42]. The reference values of densities were derived from a whole-body mouse CT scan by using a clinical Siemens Biograph CT (Siemens, Berlin, Germany) at 120 kVp and 0.6-mm slice thickness. In this scan, the mass density of the soft tissue was found to be ~ 1.02 g/cm$^3$ and the mass density for the bone was ~1.20 g/cm$^3$ based on the HU clinical calibration curve for the CT scanner used at our institution.

Notably, this scan is different from the micro-CT scan that was used to obtain the contours and the subsequent mesh files; this scan was only used to establish reference densities.

The initial printing parameters were set based on work done by Ma *et al*[41]. Printing parameters such as print flow, temperature, print speed, layer height, and others were further optimized by printing 10 $mm^3$ cubes and weighing them on a high-resolution digital scale. The printing parameters were tuned iteratively till the 10 $mm^3$ cubes yielded the required density.  Once the printing parameters were optimized (Table S3), multiple phantoms were printed with the same print settings to establish reproducibility between prints[42]. The print-to-print reproducibility of the printed phantoms was also established by delivering the same dose under identical setup conditions and comparing the dose distributions for two phantoms. The HU values and density of the 3D printed phantom was also compared with those of the reference CT scan. All irradiations in this study were repeated 5 times unless otherwise noted.

## 2.3    FLASH vs CONV Comparison

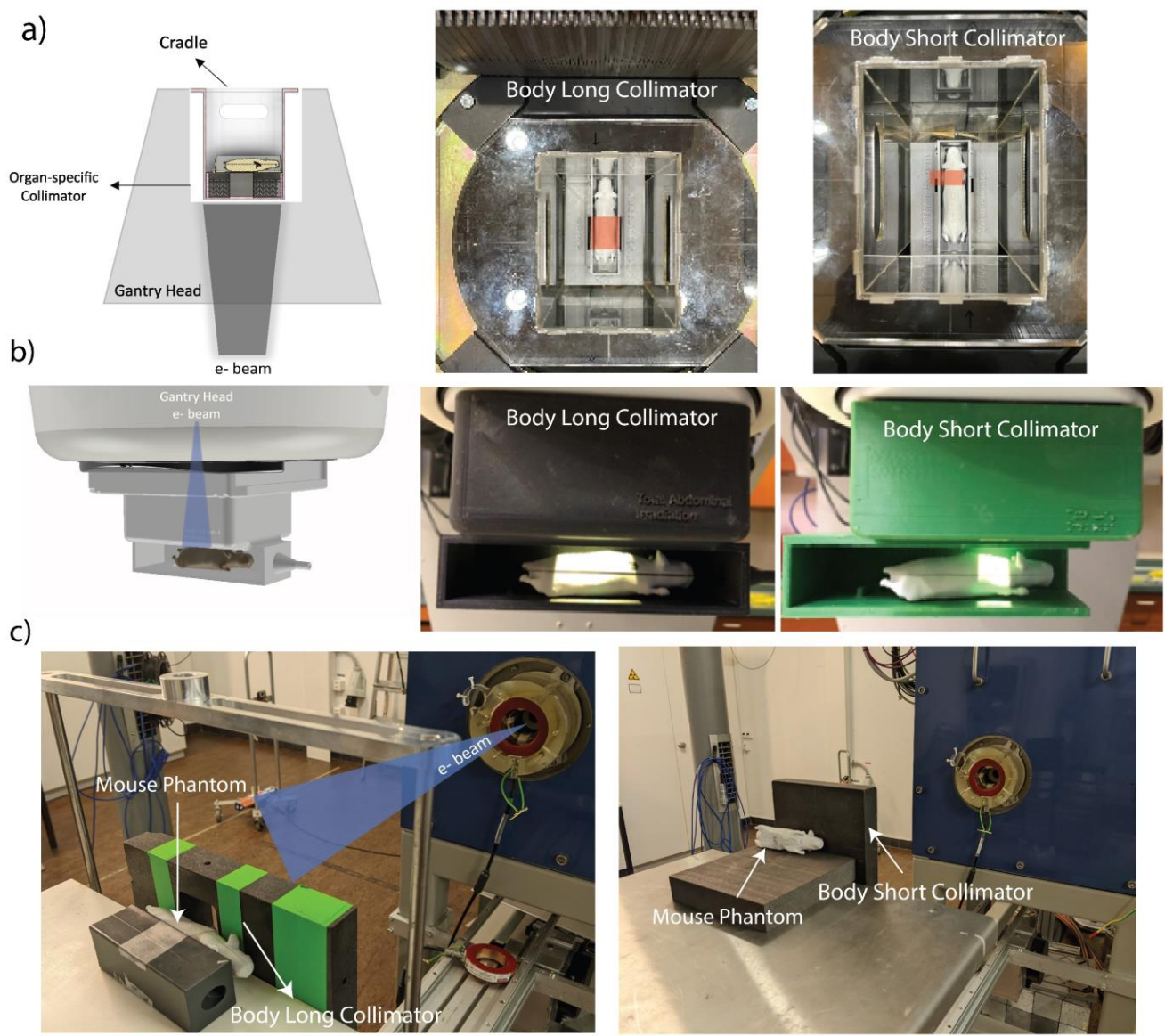

**FIGURE 2**. *Irradiation geometries for the three institutions. In panel (a), a clinical linear accelerator was configured at Stanford to deliver FLASH dose rates. The gantry cover was removed to allow the cradle to be placed near the scattering foil. The beam enters from underneath the phantom. The mouse phantom was placed in organ-specific 3D printed collimators housed inside a cradle. The Body Long and the Body Short collimators are shown here, with the red boxes indicating the irradiated area. (b) The irradiation geometry for MD Anderson. A FLASH-enabled interoperative electron linac was used. 3D-printed organ-specific collimators were attached to the gantry head. The beam enters from the top. The whole abdomen collimator is shown in the middle panel and the whole lung collimator in the left panel. (c) The irradiation setup for CHUV shows setup for Body Long and Brain irradiations.*

**TABLE 1**. Electron beam parameters at the three institutions.

| | Stanford | | MD Anderson | | CHUV | |
|---|---|---|---|---|---|---|
| | FLASH | CONV | FLASH | CONV | FLASH | CONV |
| Beam energy, MeV | 17 | 16 | 9 | 9 | 5 | 6 |
| Target dose, Gy | 10 | 10 | 10 | 10 | 10 | 10 |
| Delivered Pulses | | | | | | |
| - Body Long | 5 | ~691 | 4 | ~864 | 10, 5* | ~1600 |
| - Body Short | 5 | ~680 | 2 | ~1030 | NA | NA |
| - Brain | 5 | ~745 | 5 | ~864 | 1, 5* | ~705 |
| Pulse rate, Hz | 90 | 90 | 90 | 30 | 100 | 10 |
| Dose per pulse, Gy | | | | | | |
| - Body Long | 2 | 1.45E-2 | 2.5 | 1.16E-2 | 1, 2* | 6.25E-3 |
| - Body Short | 2 | 1.47E-2 | 5.5 | 0.97E-2 | NA | NA |
| - Brain | 2 | 1.34E-2 | 2 | 1.16E-2 | 10, 2* | 1.42E-2 |
| Dose rate, Gy/s | | | | | | |
| - Body Long | 225 | 1.30 | 300 | 0.35 | 111, 250* | 6.25E-2 |
| - Body Short | 225 | 1.32 | 900 | 0.29 | NA | NA |
| - Brain | 225 | 1.21 | 225 | 0.35 | 4.76E+6, 250* | 0.14 |
| Pulse width, µs | | | | | | |
| - Body Long | 3.75 | 3.75 | 1.2 | 1.2 | 2.1 | 1 |
| - Body Short | 3.75 | 3.75 | 4 | 1.2 | NA | NA |
| - Brain | 3.75 | 3.75 | 1 | 1.2 | 2.1 | 1 |
| Intra-pulse dose rate, Gy/s | | | | | 0.47E+6, | |
| - Body Long | 0.53E+6 | 3.86E+4 | 2.08E+6 | 9.64E+4 | 0.95E+6 | 6.25E+3 |
| - Body Short | 0.53E+6 | 3.92E+4 | 1.38E+6 | 8.09E+4 | NA | NA |
| | | | | | 4.76E+6, | |
| - Brain | 0.53E+6 | 3.58E+4 | 2.00E+6 | 9.64E+4 | 0.95E+6 | 1.42E+4 |

*Parameters for first and second replicates, respectively.

The institutions were instructed to deliver 10 Gy to the plane of film for three fields: (1) Body Long, (2) Body Short, and (3) Brain. Two replicates of the entire irradiation experiment were performed by each institution, and in each replicate, 5 films per field were irradiated under identical conditions (except for the irradiations at CHUV, in which the total dose was the same but the number of pulses and dose per pulse differed between replicates because of technical issues with the irradiator).

Essential beam parameters such as repetition rate, pulse width, beam energy and the date and time of irradiation were recorded. The three FLASH beamlines used in this study were a clinical linac configured to deliver FLASH[24,43], a commercially available FLASH-enabled Mobetron[44], and an Oriatron eRT6 electron irradiator[45]. Beam parameters such as beam energy and temporal metrics are shown in Table 1. The beam energies for the three institutions ranged from 5 MeV to 17 MeV; the irradiation geometry at the three institutions is shown in Figure 2. At Stanford (Fig. 2a), the gantry is rotated 180 degrees and the beam enters from underneath. The gantry cover was removed and the mouse phantom, which is registered to the anatomic field–specific collimator (Body Long and Body Short setups shown), was placed inside the treatment head by using a custom cradle mounted to the face of the gantry. At MD Anderson (Fig. 2b), the gantry is set at 0 degrees and the beam enters the phantom from above. The mouse phantom was aligned to the anatomic field–specific collimator (Body Long and Body Short setups shown), which is mounted to the face of the gantry. At CHUV (Fig. 2c), the system has a horizontally oriented beam at 90 degrees. The mouse phantom is positioned sideways behind the anatomic field specific collimator (Body Long and Brain setups shown). Notably, precise alignment of the fields to specific body regions was not specified or evaluated, because the different institutions used different localization methods and because spatial targeting accuracy was not the focus of this study.

## 3  Results

### 3.1 Film Calibration

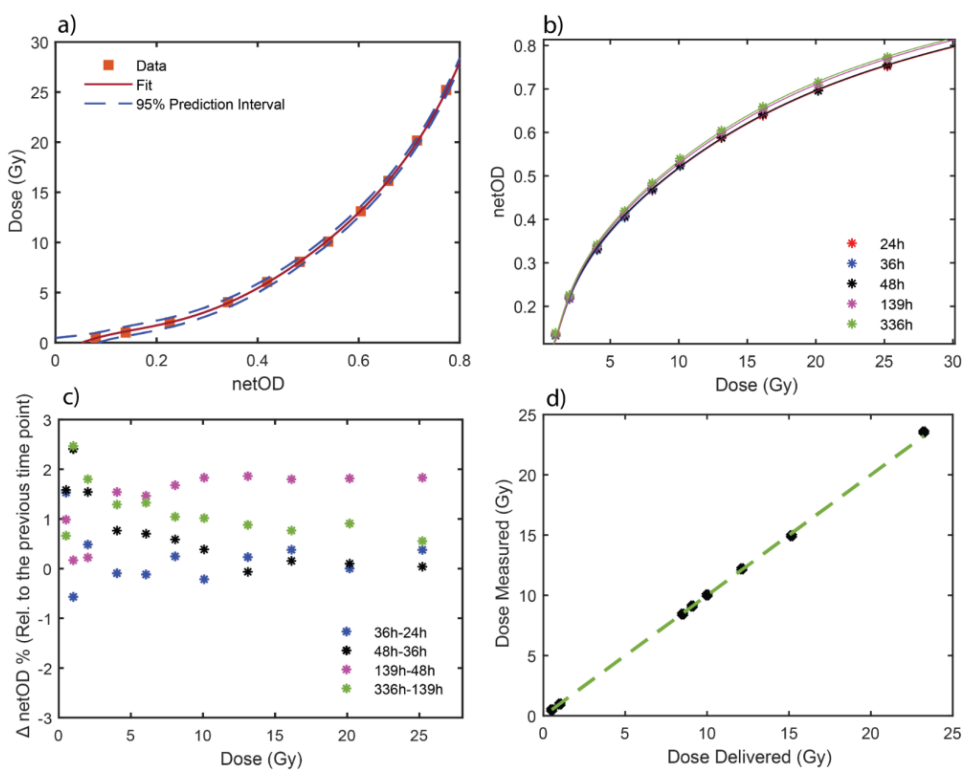

**FIGURE 3**. *Film dosimetry calibrations. (a) Film calibration curve and 95% prediction interval. A 5$^{th}$ order polynomial was used for fitting. (b) Changes in the dose response of netOD over time. (c) Percentage difference in net OD for the same dose from one time point to the next. The uncertainty associated with using a calibration curve across these different time points can be as high as ~2%. (d) Validation of the calibration curve by delivering known doses under reference conditions. The largest error between*

*measured and delivered dose was 0.32 Gy for a prescribed dose of 23 Gy (1.4%). The green dashed line*
*represents theoretical agreement for reference (slope = 1).*

The film calibration curve, with the 95% prediction interval, is shown in Figure 3a. The 95% prediction
interval was not constant with dose, which implies that the overall uncertainty will depend on the dose
level. The effect of post-irradiation changes in netOD is shown in Figure 3b; the percent change in netOD
relative to the previous time point is shown in Figure 3c. For example, the percent change in netOD
between the 24–36 h interval and the 36–48 h interval was within ±0.5%. This is a source of uncertainty
if films are not scanned close to the time-point calibration curves; indeed, errors as large as ~2.5% are
evident for at 336 h after irradiation. This finding underscores the importance of having several
calibration curves to choose from to best match the time delay between irradiation and scanning of the
experimental films[38]. A graph of delivered vs measured dose is shown in Figure 3d, where the dashed
green line serves as a visual aid and represents a slope of 1, indicating ideal agreement between
delivered and measured dose. The absolute average difference was 0.09 Gy, with the largest deviation
(0.32 Gy) seen at the 23-Gy dose level (corresponding to 1.4%).

    To quantify the overall uncertainty in the doses reported in this dosimetric audit, an uncertainty
"budget" analysis is presented in Supplementary Table S1. The uncertainty varies by the dose because of
curve fitting (as indicated by the 95% prediction interval in Fig. 3a). Typical film scanning artifacts were
also considered and as described above, were minimized as much as possible. At a dose level of 10 Gy,
we estimate the uncertainty to be ~2.1 %. The uncertainty variability in time elapsed between
irradiation and scanning was estimated from the data in Figure 3c to be ~0.5%.

## 3.2 HU Values, Mass Densities and Reproducibility

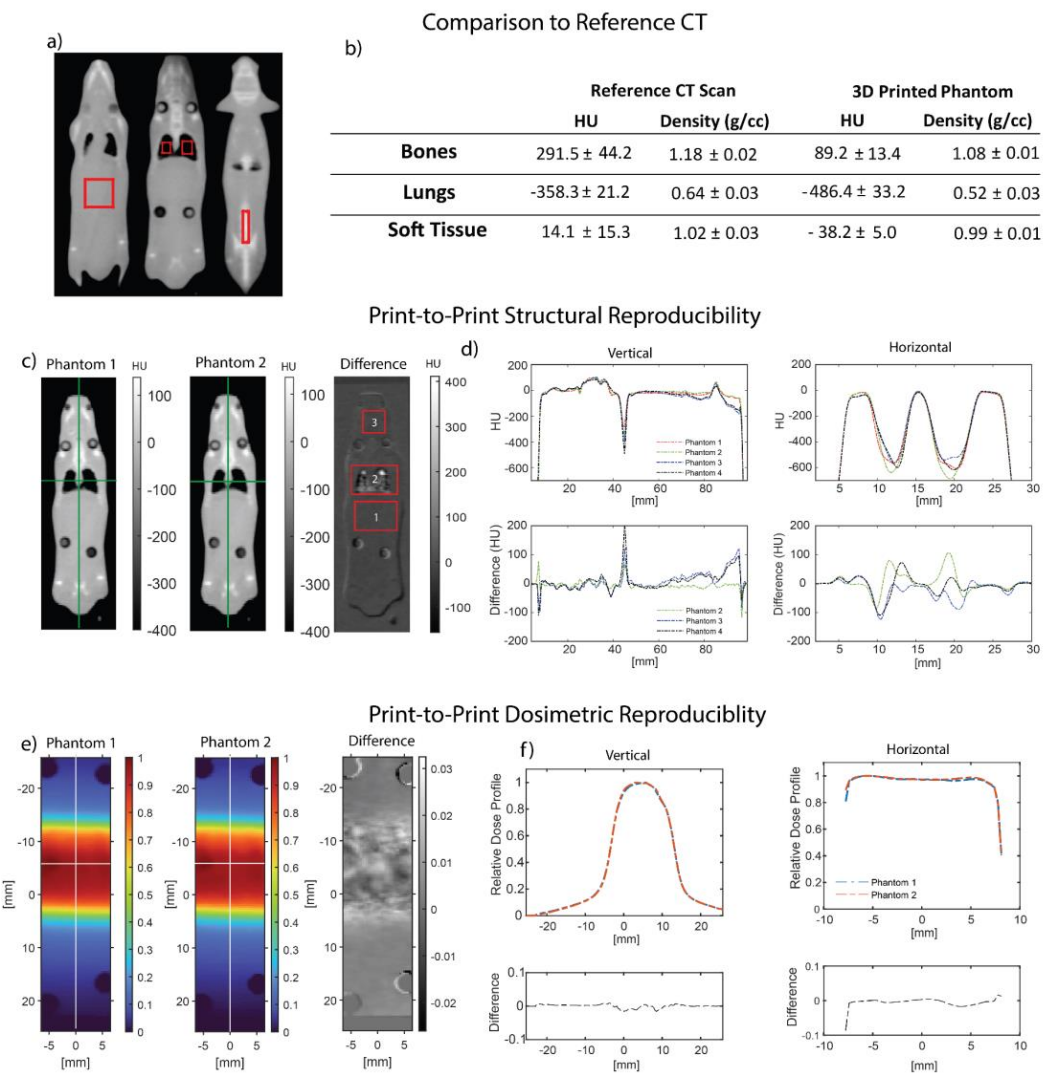

**FIGURE 4**. *Characterization of 3D mouse phantom. (a) 2D coronal slices of the CT scanned images of the mouse phantom, with region of interests (ROIs) (in red) used to compare the HU values between the reference CT of the live mouse and the 3D printed phantom. (b) Comparisons of mean HU values show small, calculated mass density differences of 0.1 g/cm³ for bone, 0.12 g/cm³ for lung, and 0.03 g/cm³ for soft tissue regions between the phantom and the live mouse. (c) 2D coronal slices of CT scans of two 3D printed phantoms. The three marked ROIs were used to quantify the difference in soft tissue (ROI1), lung (ROI2), and bone (ROI3) regions from one print to another. CT images for the 3rd and 4th phantom are shown in the Supplementary figures. The average differences in absolute HU across 4 phantoms (with phantom 1 as reference) were 8.8 HU for soft tissue, 59.7 HU for lung, and 10.7 HU for bone. (d) Line profiles for the 4 phantoms across the pixels indicated in green in (c), with differences in the profiles of the lines also shown. The average HU difference between the line profiles in (d) and (e) were 17.4 HU (d) and 21.9 HU (e) across the 4 phantoms (with phantom 1 as the reference), consistent with excellent print-to-print reproducibility. (e) Comparison of whole-lung irradiation between phantoms 1 and 2 indicates that small differences in HU values have minimal effects on dose distributions.*

Characterizations of the 3D-printed phantom in terms of its HU values, print-to-print structural variability, and dosimetric reproducibility are shown in Figure 4. In panel (a), coronal slices of the CT scan of the phantom are shown with the three ROIs (corresponding to soft tissue, lungs, and bony anatomy) used to compare the phantom to the reference CT scan of the live mouse. Measured HU values and calculated mass densities for the 3 structural anatomies are shown in panel (b) and were derived from the clinical HU-to-mass density calibration curve. Notably, the clinical CT calibration curve underestimated the density of the printed material when compared with the values obtained by physically measuring the mass density of the printed materials (Supplementary Table S2). The maximum calculated density difference of 0.12 g/cm$^3$ between the phantom and the live mouse was in the lung. Outstanding consistency between prints was evident when comparing the CT images between the 3D-printed phantom replicas (Fig. 4c and 4d; the four printed phantoms used for the reproducibility assessment are shown in Supplementary Figure S1). In Figure 4c, ROI 1 corresponds to the soft tissue, ROI 2 to the lung, and ROI 3 to the brain region. The average differences in absolute HU values across the four phantoms (with phantom 1 as reference) were 8.8 HU for the abdomen, 59.7 HU for the lung, and 10.7 HU for the brain.

The largest variation was seen in the lung (ROI 2). However, if dose distribution between the two phantoms is similar, then under similar irradiation conditions and dose prescription, any discrepancy between the phantoms can be considered negligible. This experiment was done by prescribing the same dose for the Body Short field placed over the lung region and comparing the 2D dose distribution from the films between the two phantoms. Each phantom was irradiated three times, and the 2D dose film distributions were averaged for each phantom. The results show that the average difference between the normalized dose images for phantoms 1 and 2 was 0.3% (Fig. 4e-f). The equivalent dataset for irradiations to the Brain and Body Long (placed over the abdominal region) is shown in Supplementary Figure S2. To account for the variability in the FLASH dose delivery process, the images were normalized to their respective maximum dose values. For both the Body Long and Brain collimators, the average normalized difference between the two images was ~0.5%.

## 3.3 FLASH vs CONV Dose Comparison

FLASH vs CONV 2D dose maps and 1D dose profile distribution comparisons for the Body Long, Body Short, and Brain irradiation fields for the two replicates are shown in Figure 5. Within each replicate, irradiations to each anatomic region were repeated five times. Each panel shows representative film 2D dose maps for FLASH and CONV and the difference between the dose maps. The white cross hair lines on the 2D dose maps denote the pixels where the 1D vertical and horizontal line profiles were obtained. The vertical and horizontal 1D dose profiles of both FLASH and CONV are also presented in a combined plot, and the difference between the 1D dose profiles is also plotted. The red rectangles on the 2D dose map's difference represent the ROIs used to derive the mean dose values. These mean dose values were used to determine the average dose difference between FLASH and CONV and the dose difference from the prescribed dose for each replicate; results of these comparisons are given in Table 2. The values are also presented as percentage difference from the prescribed dose. For all FLASH vs CONV comparisons, CONV was chosen as the reference distribution.

A)

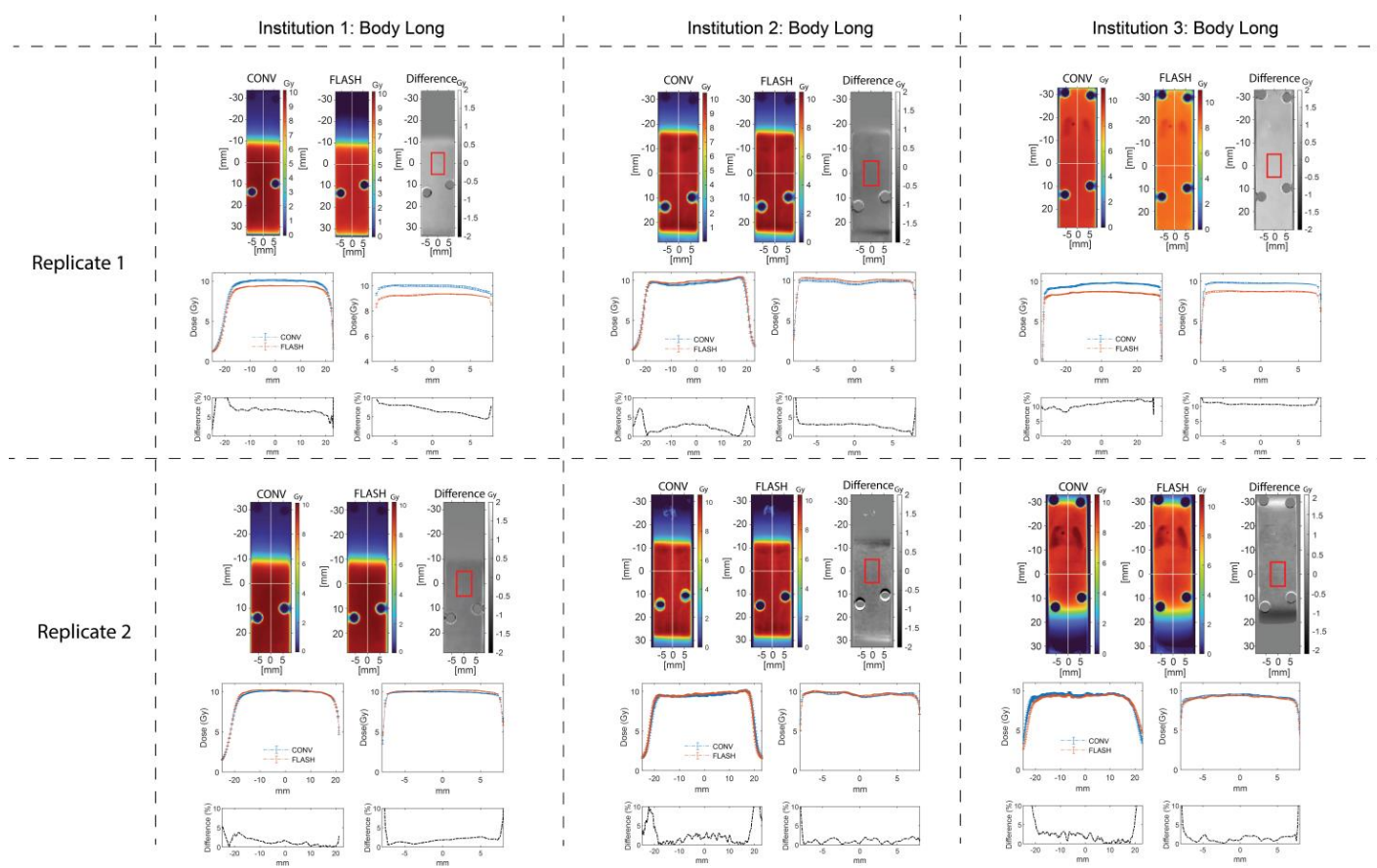

(B)

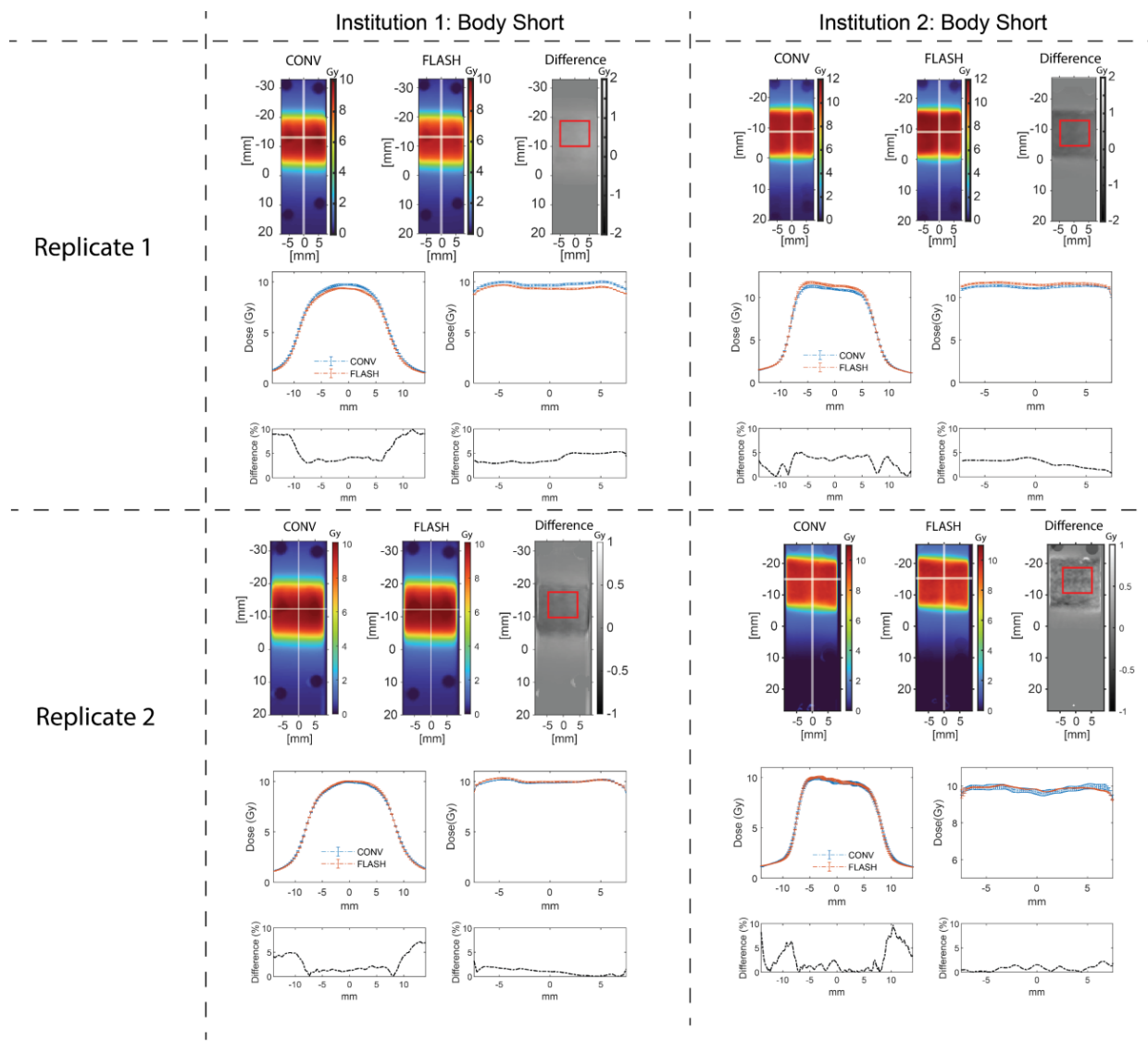

(C)

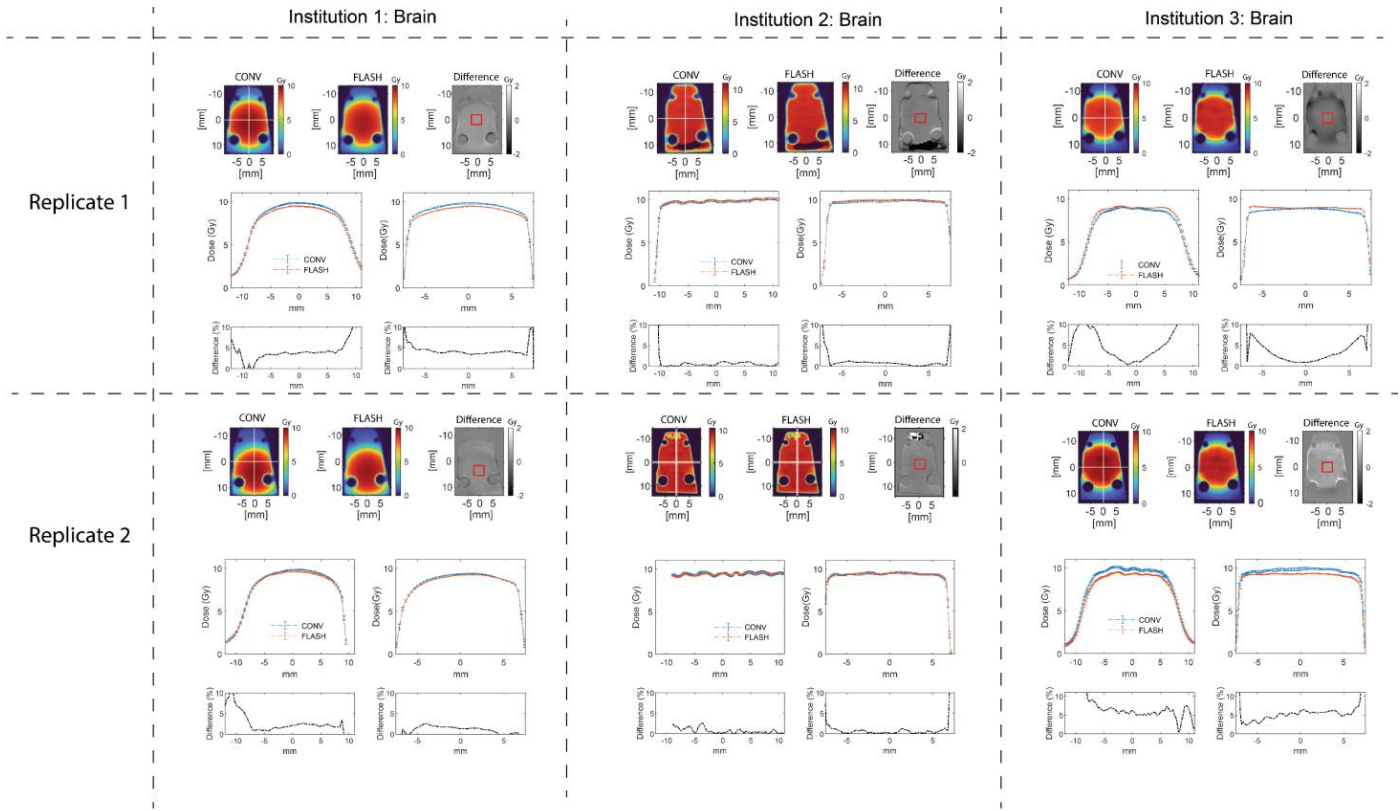

**FIGURE 5**. *2D dose comparison of FLASH vs CONV irradiations for the (A) Body Long, (B) Body Short, and (C) Brain fields for the three institutions (anonymized). Each institution performed two replicates, with 5 films irradiated per anatomic region in each replicate. For each field, the same ROIs were chosen on each film (red rectangles) to calculate the average dose difference. White lines indicate the regions that were used for line profiles. Both horizonal and vertical line profiles are shown for all irradiations. The difference between the CONV and FLASH profiles is shown under each graph. Error bars represent the standard error obtained from 5 repeated measurements. Summary statistics are shown in Table 2. Overall, discrepancies between FLASH and CONV, and discrepancies from the prescribed dose, were small across all institutions and tended to decrease between the first and second replicates.*

**TABLE 2**. Dose comparison statistics.

| Average Difference Between FLASH and CONV (for prescribed dose of 10 Gy) | | | | | |
|---|---|---|---|---|---|
| **Overall (For All Irradiation Fields) (%)** | | **Irradiation Field** | **Overall (%)** | **By Replicate (%)** | |
| 2.8 ± 2.5% | | | | | |
| | | **Body Long** | 4.0 ± 1.4% | Replicate 1 | 6.7 ± 1.5% |
| | | | | Replicate 2 | 1.3 ± 1.3% |
| **By Replicate (Gy)** | | | | | |
| Replicate 1 | 4.3 ± 2.8% | **Body Short** | 2.8 ± 2.6% | Replicate 1 | 4.3 ± 2.8% |
| Replicate 2 | 1.2 ± 0.6% | | | Replicate 2 | 1.2 ± 2.4% |
| | | **Brain** | 2.2 ± 2.1% | Replicate 1 | 2.3 ± 2.1% |
| | | | | Replicate 2 | 2.4 ± 2.4% |

| Average Difference From Prescribed Dose of 10 Gy | | | | | |
|---|---|---|---|---|---|
| **Overall (For All Irradiation Fields) (%)** | | **Irradiation Field** | **Overall (%)** | **By Replicate (%)** | |
| CONV: 3.1 ± 2.7% | | | | | |
| FLASH: 4.6 ± 2.8% | | | | | |
| | | **Body Long** | CONV: 2.5 ± 0.9% | Replicate 1 | CONV: 2.2 ± 1.1% |
| | | | FLASH: 4.5 ± 0.7% | | FLASH: 6.9 ± 0.6% |
| | | | | Replicate 2 | CONV : 2.7 ± 0.6% |
| | | | | | FLASH: 2.6 ± 0.7% |
| **By Replicate (Gy)** | | | | | |
| Replicate 1 | CONV: 3.6 ± 3.8% | **Body Short** | CONV: 0.7 ± 1.4% | Replicate 1 | CONV : 2.0 ± 1.5% |
| | FLASH: 6.4 ± 3.9% | | FLASH: 3.4 ± 1.3% | | FLASH: 6.0 ± 1.4% |
| Replicate 2 | CONV: 2.5 ± 2.0% | | | Replicate 2 | CONV : 1.4 ± 1.2% |
| | FLASH: 2.7 ± 1.7% | | | | FLASH: 0.7 ± 1.2% |
| | | **Brain** | CONV: 1.9 ± 2.1% | Replicate 1 | CONV: 6.0 ± 2.4% |
| | | | FLASH: 5.6 ± 0.9% | | FLASH: 6.1 ± 1.2% |
| | | | | Replicate 2 | CONV: 2.9 ± 0.8% |
| | | | | | FLASH: 5.1 ± 0.6% |

Overall, across all institutions, replicates, and irradiation fields, for a prescribed dose of 10 Gy, the average dose difference between FLASH and CONV was 0.28 ± 0.25 Gy (2.8% ± 2.5%), and the average dose difference from the prescribed dose was 0.46 ± 0.28 Gy (4.6% ± 2.8%) for FLASH and 0.31 ± 0.27 Gy (3.1% ± 2.7%) for CONV. Discrepancies seemed to decrease from the first replicate to the second, with the mean difference between FLASH and CONV decreasing from 4.3% to 1.2%, and the mean difference from the prescribed dose decreasing from 6.4% to 2.7% for FLASH and 3.6% to 2.5% for CONV. On a per-institution basis, the maximum average difference between FLASH and CONV was 1.02 Gy (10.5%) and the maximum average difference from the prescribed dose was 1.34 Gy (13.4%) in the first replicate, decreasing in the second replicate to 0.40 Gy (3.6%) and 0.53 Gy (5.3%), respectively. Data for each institution and each individual irradiation field are shown in Supplementary Table S4.

# 4  Discussion

In this study, we developed, optimized, and 3D-printed an anatomically realistic mouse phantom for use in multi-institutional dosimetric audits of FLASH and CONV dose distributions. We chose these materials because their inherent densities closely match the densities in the reference CT scan of the live mouse. Typically, varying density is achieved by varying the infill percentage of the prints[46,47]. However, the infill pattern can affect scattering of photons and electrons and can also produce different dose distributions depending on where the beam is incident on the phantom[48]. Therefore, the optimal method to vary density is to choose materials that inherently have different densities and use the maximum infill percentage[41]. In addition to ABS and PLA, a third material, lightweight PLA, was used to print lungs. The extruded volume of lightweight PLA can be controlled by tuning the printing temperature, speed, and flow. At about 230°C, lightweight PLA starts to expand and 'foams up'. Densities as low as 35% of typical PLA can be achieved if appropriate settings are used. The final printed densities for the three materials and the optimized printing parameters are shown in Supplementary Tables S2 and S3.

In future iterations, we intend to use 4 or more printing materials simultaneously by using an automatic material exchange station. In the current study, we established excellent print-to-print reproducibility, as demonstrated by comparisons of HU profiles, irradiating under same geometry, and comparing the 2D dose distributions. Excellent agreement was seen (<0.5% difference in dose distributions) between the phantoms. The HU values were also in agreement (<59.7 HU difference), with the largest variation in the lung region. These values represent tighter consistency than has been reported in previous studies; Craft *et al* found 7% variability in density of repeated prints of PLA and 3% variability in density of repeated prints of ABS[42]. This emphasizes the importance of high-quality printing. Two potential reasons for the greater variation in lungs HU values could be because (1) the 0.4-mm nozzle could not completely reproduce the lung contour with fine details, meaning that some edges of the 3D-printed lung had to be manually trimmed for it to fit in the cavity; and (2) because lightweight PLA expands, complex shapes will necessarily be somewhat different from the original 3D design file. At first glance, the HU values do not seem to represent the reference CT scan (Fig. 4b); however, the clinical HU-to-electron-density calibration curve for most printing materials is typically not valid, as noted by Craft *et al*[42]. That group proposed scalar factors, 1.058 for PLA density and 1.038 for ABS density, if they were to be modeled accurately by the clinical HU-to-density calibration curve. We observed a similar trend in the current study, in which the density correction factors for PLA and ABS were 1.092 and 1.030 if they were to be accurately modeled by the clinical HU-to-density calibration curve. This may be because the exact chemical composition of commercially available printing materials is not well characterized. The CT calibration curve would underestimate mass density, and physically measuring mass density is a more appropriate method of determining density. The physical densities of printed cubes of each material (Suppl. Table S2) and the in-situ density of these materials within the phantom as determined by CT scan (Fig. 4b) demonstrated overall good agreement with the values measured on the reference CT scan of the live mouse. Notably, small deviations in the HU or density will not considerably change the delivered dose for electron beams; Fang *et al*[49] concluded that reasonable deviations in the HU-to-density curve do not meaningfully influence target dose, but doses at deeper depths can be affected. Another consideration is whether this phantom would be appropriate for other FLASH modalities such as protons and kV photons. In the current study, only material density was optimized and matched to realistic anatomic densities; the material composition (Z) was not considered, especially for bones. This would have a minimal effect on electron beams, as the collisional stopping power is independent of energy. However, for kV photons, differences in material composition can result in large differences in dose deposition, especially for bone. Moreover, for protons, the important point is to match the stopping power ratio of a material for it to be a good tissue surrogate. Tissue surrogates routinely used for photon end-to-end testing will not be valid for protons[50]. Therefore, the phantom in its current form cannot be considered suitable for dosimetric audits for protons and kV photon FLASH.

A multi-institutional audit of FLASH vs CONV can be used to answer three important questions: (1) how close are the dose and the dose distribution between FLASH and CONV? (2) how accurate was the dose delivery relative to the prescribed dose? and (3) how accurately can FLASH and CONV doses to specific anatomic regions be delivered between institutions? In this study, the first two questions were addressed. However, the institutions were instructed only to deliver a specified dose at the plane of the film; we did not specify any immobilization method or use of image guidance. A more comprehensive evaluation would also include assessing the targeting accuracy for different anatomic regions, which would then answer the third question raised above. Moreover, the phantom reports dose only on a single plane. Percentage depth dose differences will likely cause the integral dose to be different. In future iterations, a vertically cut phantom will be used to assess the integral dose as well.

 The average of the two replicates across all irradiation fields suggests that all institutions were able to deliver FLASH and CONV doses that agreed to within 3% of each other and 5% of the prescription dose. This interesting finding suggests that the dose uncertainty stems largely from lack of treatment planning options, as evidenced by the larger difference from prescribed dose relative to the difference between the modalities.

Similar dosimetric audits have been conducted in the past for clinical trials, and those audits served as credentialing processes that taught centers how to prescribe and deliver doses adequately. For centers looking to perform similar audits for FLASH, the mouse phantom mesh files are available athttps://drive.google.com/drive/folders/1N43SPe-oL3njeNjjITyG2v_nogTprsjE and the print settings and materials used are shown in Supplementary Tables S2 and S3.

Another notable point was that the institutions that participated in the study have substantial experience with FLASH delivery. Overall, the consistency of delivered doses with the prescribed dose was high in both FLASH and CONV, yet the apparent improvement between replicates suggests that familiarity with a given experimental setup may have an impact. Presumably a broader audit of institutions conducting preclinical FLASH research would have larger discrepancies, at least initially. This highlights the importance of the dosimetric tools and methods we have introduced here, which could serve as a framework for both training and credentialing for FLASH preclinical research, and later, for clinical trials[26].

## 5  Conclusion

Addressing the critical need for reproducibility in preclinical research requires that dosimetric consensus across institutions be attained. To achieve this, dosimetry metrics must be verified independently for institutions performing FLASH irradiations. In this study, we developed a 3D-printed, anatomically realistic mouse phantom to be used in a dosimetric audit. To the best of our knowledge this is the first instance of a dosimetric audit for FLASH while using realistic geometry. Although these results are accurate relative to past preclinical dosimetric audits, there is still room for improvement as the margin of error is small for FLASH. A follow-up study will include more institutions, assessment of anatomic targeting accuracy, assessment of depth dose with a vertically cut mouse phantom, and additional dosimeters. Nevertheless, the framework presented here is promising for credentialing multi-institutional studies of FLASH preclinical research to maximize the reproducibility of biological findings.

*Supplementary Table S1: Film Uncertainty Analysis*

| | Sources of Uncertainty | | Rel. Standard Uncertainties (k=1; coverage factor) |
|---|---|---|---|
| 1) | Absolute Absorbed Dose (TG 51) Co$^{60}$ $N_{d,w}$ | | 1.1 % |
| 2) | Calibration of film (Experimental) | a) | Profile flatness within 1% |
| |   a) Films placed side by side (i.e., effect of dose profile) | b) | 0.1%* |
| |   b) Scanning Orientation | c) | 0.1 %* |
| |   c) Scanner Uniformity | d) | 0.5 % |
| |   d) Variability in time elapsed between irradiation and scanning | | |
| 3) | Calibration of film (Fitting) (10 Gy) | | 1.1% |
| | Total Combined Uncertainty | | 2.1 % |

\* Sorriaux, J., Kacperek, A., Rossomme, S., Lee, J.A., Bertrand, D., Vynckier, S., Sterpin, E., 2013. Evaluation of Gafchromic® EBT3 films characteristics in therapy photon, electron and proton beams. Physica Medica 29, 599–606. https://doi.org/10.1016/j.ejmp.2012.10.001

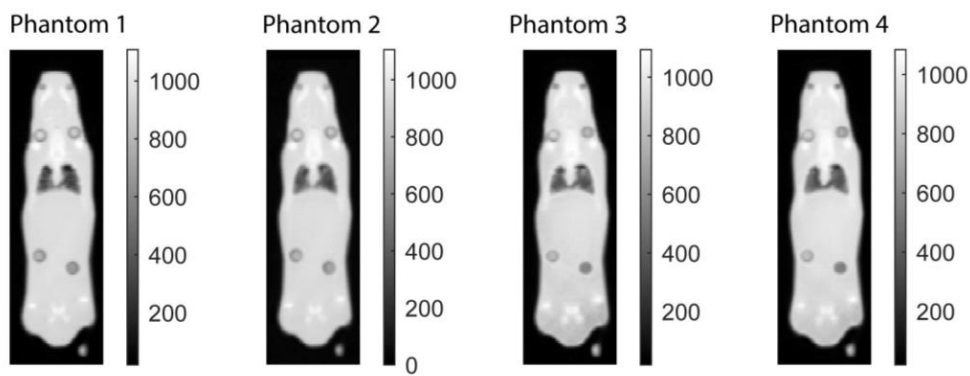

*Supplementary Figure S1. Coronal CT slices for 4 printed mouse phantoms. The vertical and horizontal line profiles for these phantoms are shown in Figure 4d.*

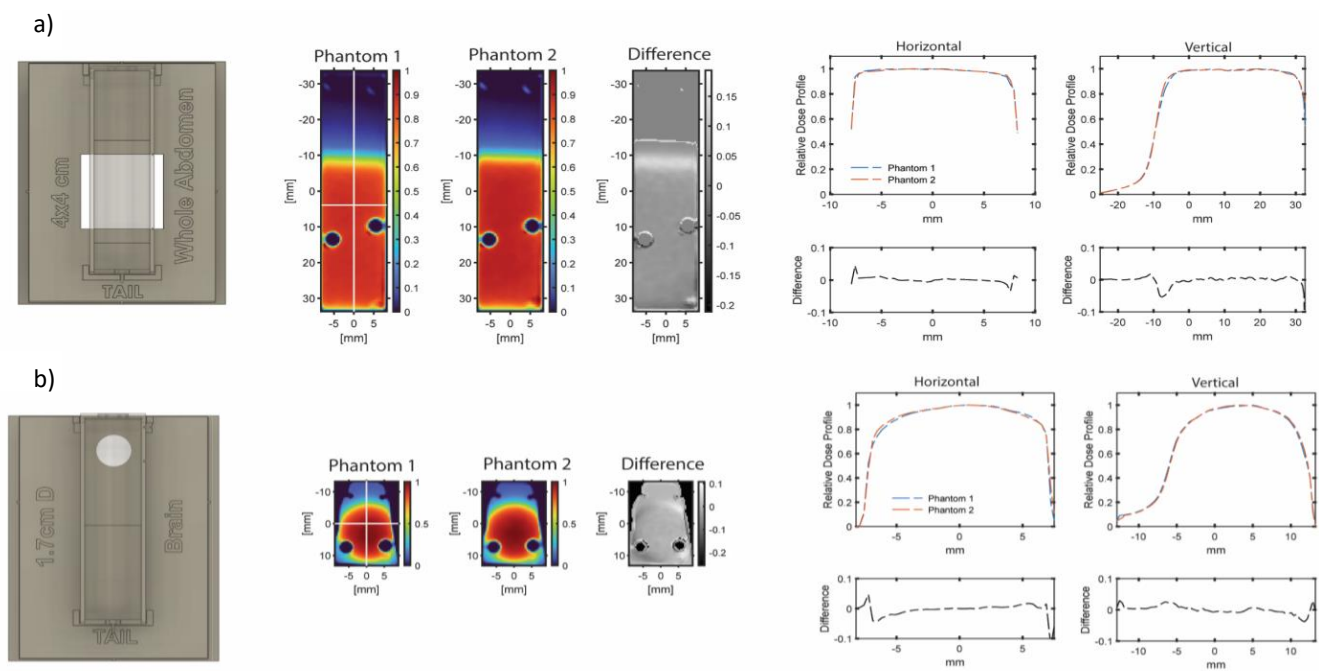

*Supplementary Figure S2. (a) Data for the Body Long irradiation for the two phantoms at FLASH dose rates. 2D images and 1D line profiles are displayed. For the line profiles, the graph on the left is for the horizontal line profile, whereas the graph on the right is for the vertical line profile. (b) The same data for the Brain irradiation. The average differences between the normalized images for the Body Long and Brain irradiation were within 0.5%.*

*Supplementary Table S2: Mass densities of the 3D printed cubes for the three printing materials used in this study*

| Organ | Printing Material | Dimensions (mm) | Weight (g) | Density ($\frac{g}{cc}$) |
|---|---|---|---|---|
| Soft Tissue | ABS | 10.3 x 10.2 x 10 | 1.06 | 1.01 |
| Spine | PLA | 9.8 x 10.2 x 10.1 | 1.24 | 1.22 |
| Lung | LW PLA | 10.1 x 10.1 x 10.1 | 0.45 | 0.44 |

*Supplementary Table S3: Optimized printing parameters*

|  | ABS | PLA | LW PLA |
|---|---|---|---|
| Line Width | 0.4 mm | 0.4 mm | 0.4 mm |
| Infill Density | 100% | 100% | 100% |
| Flow | 100% | 110% | 35% |
| Print Speed | 60 mm/s | 50 mm/s | 45 mm/s |
| Infill Speed | 60 mm/s | 50 mm/s | 45 mm/s |
| Layer Height | 0.06 mm | 0.06 mm | 0.2 mm |
| Bed Temp | 80 C | 80 C | 0 C |
| Print Temp | 250 C | 210 C | 270 C |

Abbreviations: ABS, acrylonitrile butadiene styrene; PLA, polylactic acid; LW PLA, lightweight polylactic acid.

*Supplementary Table S4: Breakdown of dose comparison by institution (anonymized) and irradiation field*

| Rx= 10 Gy | Average Dose Difference Between FLASH and CONV (Gy) | | | | | | Average Difference from Prescribed Dose (Gy) | | | | | |
| --- | --- | --- | --- | --- | --- | --- | --- | --- | --- | --- | --- | --- |
| | Institution 1 | | Institution 2 | | Institution 3 | | Institution 1 | | Institution 2 | | Institution 3 | |
| | Replicate 1 | Replicate 2 | Replicate 1 | Replicate 2 | Replicate 1 | Replicate 2 | Replicate 1 | Replicate 2 | Replicate 1 | Replicate 2 | Replicate 1 | Replicate 2 |
| **Body Long** | -0.68 ± 0.20 (6.8%) | -0.18 ± 0.07 (1.8%) | -0.27± 0.12 (2.8 %) | -0.09 ± 0.16 (1.0%) | 1.02 ± 0.13 (10.5%) | 0.11 ± 0.17 (1.1%) | (CONV) -0.07 ± 0.16 (0.7%) (FLASH) -0.75 ± 0.04 (7.5%) | (CONV) -0.11±0.04 (1.1%) (FLASH) 0.07 ± 0.03 (0.7%) | (CONV) -0.32 ± 0.08 (3.2%) (FLASH) -0.05 ± 0.04 (0.4%) | (CONV) -0.48 ± 0.08 (4.8%) (FLASH) -0.38 ± 0.08 (3.8%) | (CONV) -0.26 ± 0.09 (2.6%) (FLASH) -1.28 ± 0.11 (13.2%) | (CONV) -0.22 ± 0.06 (2.2%) (FLASH) -0.33 ± 0.11 (3.3%) |
| **Body Short** (Rx= 11 Gy Replicate 1/ Institution 2) | -0.40 ± 0.27 (4.1%) | -0.12 ± 0.16 (1.2%) | 0.40 ± 0.29 (3.6%) | -0.04 ± 0.31 (0.4%) | NA | NA | (CONV) -0.26 ± 0.20 (2.6 %) (FLASH) -0.66 ± 0.10 (6.6 %) | (CONV) -0.11 ± 0.05 (1.1 %) (FLASH) 0.01 ± 0.11 (0.1 %) | (CONV) 0.14 ± 0.11 (1.3%) (FLASH) 0.54 ± 0.18 (4.9%) | (CONV) -0.16 ± 0.19 (1.6%) (FLASH) -0.12 ± 0.12 (1.2 %) | NA | NA |
| **Brain** | -0.34 ± 0.25 (3.5%) | -0.20 ± 0.26 (2.1%) | 0.08 ± 0.34 (0.8%) | 0.04 ± 0.17 (0.4%) | 0.23 ± 0.03 (2.6%) | -0.48 ± 0.16 (4.8%) | (CONV) -0.25 ± 0.12 (2.5%) (FLASH) -0.59 ± 0.13 (5.9%) | (CONV) -0.20 ± 0.04 (2.0%) (FLASH) -0.41 ± 0.06 (4.1 %) | (CONV) -0.22 ± 0.21 (2.2%) (FLASH) -0.14 ± 0.13 (1.4%) | (CONV) -0.53 ± 0.12 (5.3%) (FLASH) -0.49 ± 0.05 (4.9%) | (CONV) -1.34 ± 0.39 (13.4%) (FLASH) -1.11 ± 0.09 (11.1%) | (CONV) -0.15 ± 0.08 (1.5%) (FLASH) -0.62 ± 0.08 (6.2%) |